\documentclass{iopart}
\usepackage{bm,graphicx,epsf}

\begin{document}

\title[Population transfer in a five-level system]{Coherently-assisted incoherent population transfer in a five-level system}

\author{S. Rebi\'{c}, D. Vitali, C. Ottaviani and P. Tombesi}
\address{Dipartimento di Fisica, Universit\`{a} di Camerino, I-62032 Camerino, Italy}

\begin{abstract}
A novel effect of population transfer in a five-level system is analyzed. This population transfer effect is found to be a version of a Raman process, which is facilitated and assisted by coherence effects, acting to close other available decay channels.
\end{abstract}

\submitto{\JPB}



\section{Introduction \label{sec:intro}}

Population transfer between individual discrete atomic states is a topic that has attracted a lot of attention in recent years~\cite{Arimondo96,Bergmann98,Vitanov01}. Usually, population transfer techniques rely on specifically tailored time dependent pulses~\cite{Bergmann98}, or on the existence of quantum coherence between the two states involved~\cite{Arimondo96}. In this article, we propose a novel method of population transfer between specified atomic ground states in a five-level system in a M configuration  scheme~\cite{Greentree03,Matsko03a,Ottaviani03,Matsko03b} (see \fref{fig:mscheme}). The method is well-suited for (although not limited to) the implementation with Rydberg atoms.

The big advantage of this technique is that it concerns steady-state populations, which means that it can be easily realized by keeping the system at its steady state and adiabatically changing the atomic detuning. Hence, there is no need for specifically tailored light pulses. Since it concerns steady-state populations, the process is also fully reversible. The present transfer mechanism between two different ground state sublevels is similar to a Raman process because it is mediated by the spontaneous emission from an excited level. However, atomic coherence also plays an important role here because it inhibits the decay into other levels. Another difference of the present population transfer effect with respect to the others ~\cite{Arimondo96,Bergmann98} is that the latter can be described by means of the amplitude equations formalism. We shall show that in the present case this formalism is insufficient and that a description in terms of Bloch equations is needed to account for this effect.

The article is organized as follows. In \sref{sec:system} we introduce the M-scheme and write the full set of Bloch equations. In \sref{sec:poptrans}, the population transfer is analyzed in detail, and alternative configurations are proposed. \Sref{sec:conclusion} outlines the conclusions.

\section{The system and the Bloch equations \label{sec:system}}

\begin{figure}
\epsfxsize=8cm
\epsfysize=5cm
\begin{center}
\epsfbox{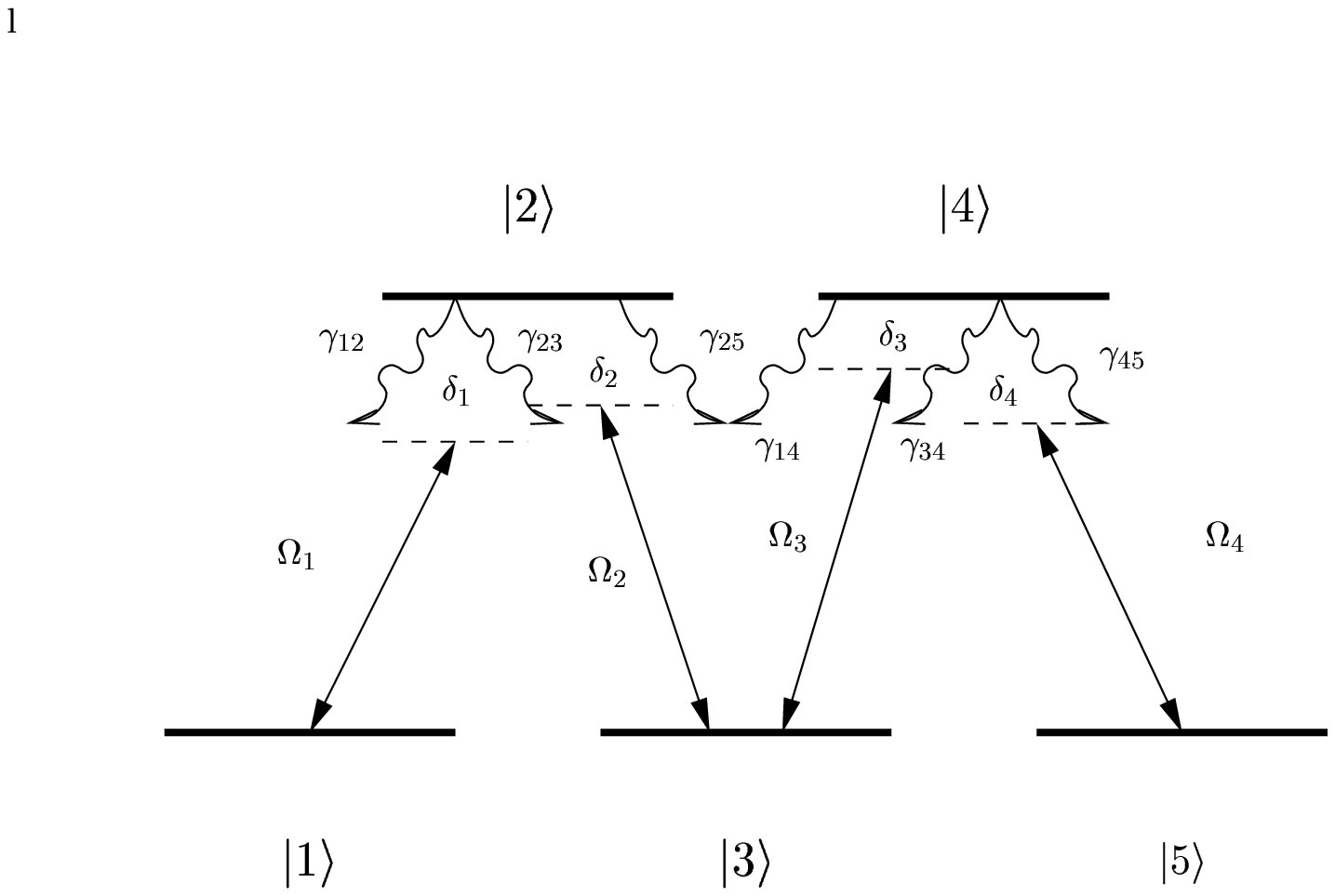}
\end{center}
\caption{\label{fig:mscheme}Schematic depiction of the M-scheme. Rabi frequencies of the optical fields driving the respective transitions are $\Omega_j$, detunings of atomic transitions and optical frequencies are $\delta_j$ and $\gamma_{ij}$ are the spontaneous emission rates in $|i\rangle \leftrightarrow |ket\rangle$ decay channel.}
\end{figure}

The system under consideration is called the M-scheme~\cite{Ottaviani03} and is shown in \fref{fig:mscheme}. The Hamiltonian of this system, in the interaction picture and rotating wave approximation, is
\begin{eqnarray}\label{eq:hambare}
H &=& \hbar\delta_{1} |2\rangle\langle2| + \hbar\delta_{12}|3\rangle\langle3| + \hbar\delta_{13}|4\rangle\langle4| + \hbar\delta_{14}|5\rangle\langle5| + \nonumber\\
&\ & + \hbar\left(\Omega_{1}|2\rangle\langle1| + \Omega_{2}|2\rangle\langle3| + \Omega_{3}|4\rangle\langle3| + \Omega_{4}|4\rangle\langle5|+h.c.\right),
\end{eqnarray}
where $\delta_{12}=\delta_1-\delta_2$, $\delta_{13} = \delta_{12}+\delta_3$ and $\delta_{14}=\delta_{13}-\delta_4$. Rabi frequencies are defined in terms of dipole matrix elements $\mu_{ij}$ between atomic levels $|i\rangle$ and $|j\rangle$, and of the amplitudes of the electric field applied to this transition $\mathcal{E}_k$, as $\Omega_k = -\mu_{ij} \mathcal{E}_k/\hbar$. 

Including decay from the excited states $|2\rangle$ and $|4\rangle $ onto the ground state sublevels $|1\rangle $, $|3\rangle $, $|5\rangle$, one gets the following set of Bloch equations for the diagonal elements of density matrix (atomic populations) 
\numparts
\begin{eqnarray}
\dot{\rho}_{11} &=& \gamma_{12}\rho_{22} + \gamma_{14}\rho_{14} + i\left(\Omega_1\rho_{21}-\Omega_1^*\rho_{12} \right) , \\
\dot{\rho}_{22} &=& -(\gamma_{12}+\gamma_{23}+\gamma_{25})\rho_{22} - i\left(\Omega_1\rho_{21}-\Omega_1^*\rho_{12} \right) \nonumber \\
&\ &-i\left( \Omega_2\rho_{23}-\Omega_2^*\rho_{32}\right), \label{eq:rho22} \\
\dot{\rho}_{33} &=& \gamma_{23}\rho_{22}+\gamma_{34}\rho_{44} + i\left( \Omega_2\rho_{23}-\Omega_2^*\rho_{32}\right) + i\left(\Omega_3\rho_{43}-\Omega_3^*\rho_{34} \right), \\
\dot{\rho}_{44} &=& -(\gamma_{14}+\gamma_{34}+\gamma_{45})\rho_{44} - i\left(\Omega_3\rho_{43}-\Omega_3^*\rho_{34} \right) \nonumber \\
&\ &- i\left(\Omega_4\rho_{45}-\Omega_4^*\rho_{54} \right) , \\
\dot{\rho}_{55} &=& \gamma_{25}\rho_{22}+\gamma_{45}\rho_{44} + i\left(\Omega_4\rho_{45}-\Omega_4^*\rho_{54} \right) , \label{eq:rho55}
\end{eqnarray}
\endnumparts
while those for atomic coherences are
\numparts
\begin{eqnarray}
\dot{\rho}_{12} &=& -\left( \gamma_d+\frac{\gamma_{12}+\gamma_{23}+\gamma_{25}}{2} + \delta_1 \right) \rho_{12} \nonumber \\
&\ &+ i\Omega_1(\rho_{22}-\rho_{11}) + i\Omega_2\rho_{13} , \\
\dot{\rho}_{13} &=& -\left(\gamma_d + i\delta_{12} \right) \rho_{13} + \Omega_1\rho_{23}-i\Omega_2^*\rho_{12} - i\Omega_3^*\rho_{14} , \\
\dot{\rho}_{14} &=& -\left(\gamma_d + \frac{\gamma_{14}+\gamma_{34}+\gamma_{45}}{2}+ \delta_{13} \right) \rho_{14} \nonumber \\
&\ &+ i\Omega_1\rho_{24}-i\Omega_3\rho_{13} - i\Omega_4\rho_{15} , \\
\dot{\rho}_{15} &=& -\left(\gamma_d + i\delta_{14} \right) \rho_{15} + \Omega_1\rho_{25}-i\Omega_4^*\rho_{14} , \\
\dot{\rho}_{23} &=& -\left(\gamma_d + \frac{\gamma_{12}+\gamma_{23}+\gamma_{25}}{2}+ \delta_2 \right) \rho_{23} \nonumber \\
&\ &+ i\Omega_1^*\rho_{13}+ i\Omega_2(\rho_{33}-\rho_{22}) -i\Omega_3^*\rho_{23} , \\
\dot{\rho}_{24} &=& -\left( \gamma_d + \frac{\gamma_{12}+\gamma_{14}+\gamma_{23}+\gamma_{25}+\gamma_{34}+\gamma_{45}}{2} - \delta_{23} \right)\rho_{24} \nonumber \\
&\ &+ i\Omega_1^*\rho_{14} + i\Omega_2^*\rho_{34} - i\Omega_3\rho_{23} - i\Omega_4\rho_{25}  \\
\dot{\rho}_{25} &=& -\left( \gamma_d+\frac{\gamma_{12}+\gamma_{23}+\gamma_{25}}{2} + i\delta_{24} \right) \rho_{25} \nonumber \\
&\ &+i\Omega_1^*\rho_{15} + i\Omega_2^*\rho_{35} -i\Omega_4^*\rho_{24} , \\
\dot{\rho}_{34} &=& -\left(\gamma_d + \frac{\gamma_{14}+\gamma_{34}+\gamma_{45}}{2}+ i\delta_3 \right) \rho_{34} \nonumber \\
&\ &+i\Omega_2\rho_{24} + i\Omega_3*(\rho_{44}-\rho_{33}) -i\Omega_4\rho_{35} , \\
\dot{\rho}_{35} &=& -\left(\gamma_d + i\delta_{34} \right) \rho_{35} + i\Omega_2\rho_{25} + i\Omega_3 \rho_{45} -i\Omega_4^* \rho_{34} , \\
\dot{\rho}_{45} &=& -\left(\gamma_d + \frac{\gamma_{14}+\gamma_{34}+\gamma_{45}}{2}+ i\delta_4 \right) \rho_{45} \nonumber \\
&\ &+i\Omega_3^*\rho_{35} + i\Omega_4^*(\rho_{55}-\rho_{44}) .
\end{eqnarray}
\endnumparts
Here, $\delta_{23}=\delta_2-\delta_3$, $\delta_{24}=\delta_{23}+\delta_4$ and $\gamma_d$ denotes the dephasing rate of the ground states, taken to be equal.

\section{The Population Transfer \label{sec:poptrans}}

\begin{figure}
\epsfxsize=12cm
\begin{center}
\epsfbox{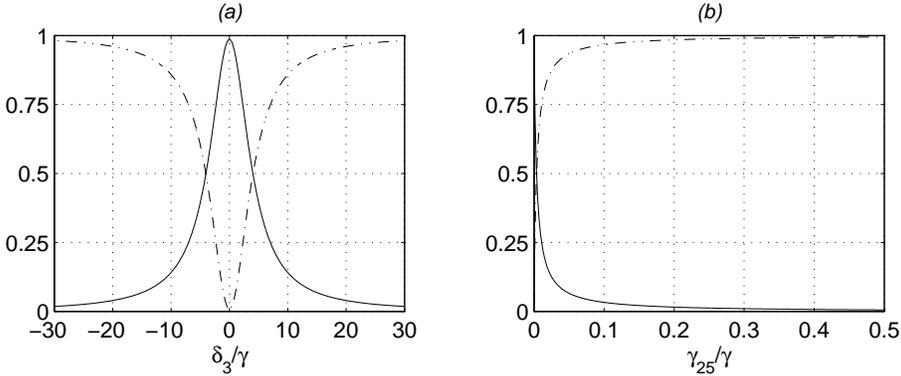}
\end{center}
\caption{\label{fig:forpub1}$(a)$ Populations $\rho_{11}$ (\full) and $\rho_{55}$ (\chain) for the following parameters: $\gamma_{12} = \gamma_{23} = \gamma_{34} = \gamma_{45} = \gamma$, $\gamma_d = 10^{-2}\gamma$, $\delta_1 = 20\gamma$, $\delta_2 = 0$, $\delta_4 = \delta_3$, $\Omega_1 = 0.75\gamma$, $\Omega_2 = 1.5\gamma$, $\Omega_3 = 0.01\gamma$ and $\Omega_4 = 0.1\gamma$, $\gamma_{14} = \gamma_{25} = 0.25\gamma$. In $(b)$, the same populations are shown for increasing $\gamma_{25}$ and fixed $\delta_3 = \delta_4 = 20\gamma$. Other parameters are the same as in $(a)$.}
\end{figure}

The M-scheme of \fref{fig:mscheme} is often used for studies related with electromagnetically-induced transparency (EIT) \cite{Arimondo96,Harris97} (see for example \cite{Greentree03,Matsko03a,Ottaviani03,Matsko03b}), and in such cases one usually assumes that the atomic population is initially concentrated in level $|1\rangle$, and that the $|1\rangle \leftrightarrow |2\rangle$ transition is excited by a probe field, weaker than the intense pump field coupling the $|3\rangle \leftrightarrow |2\rangle$ transition, i.e., $\Omega_1 < \Omega_2$. In this situation, the population typically remains trapped in state $|1\rangle$. If pump and probe fields are at two photon resonance, i.e., $\delta_1 \simeq \delta_2$, one has EIT, the dark state involving the coherent superposition of $|1\rangle$ and $|3\rangle$ is realized and the population is partially transferred to level $|3\rangle$, even though this transfer is negligible when $\Omega_1 \ll \Omega_2$. Outside EIT, for example when $\delta_1 \gg \delta_2$, one expects that the atoms do not leave state $|1\rangle$. However, when $\delta_3 \simeq \delta_4 \neq 0$ and $\Omega_3$ and $\Omega_4$ are smaller than $\Omega_1$ and $\Omega_2$, an unexpected new steady state with almost all atoms in level $|5\rangle$ becomes possible, allowing to transfer the atomic population from $|1\rangle $ to $|5\rangle$, by adiabatically sweeping the detuning $\delta_3$. In fact, when $\delta_3 \approx \delta_4$ one has two-photon resonance and therefore EIT effect in the second $\Lambda$ subsystem $|3\rangle - |4\rangle - |5\rangle$. EIT ensures that the excited state $|4\rangle$ is unpopulated and hence that the spontaneous emission decays from it can be neglected. Spontaneous emission from $|2\rangle$ instead cannot be neglected, yielding in general decay to levels $|3\rangle$ and $|5\rangle$. However, when $\delta_1 \gg \delta_2 \simeq 0$, quantum interference inhibits populating state $|3\rangle$, because the probe is highly detuned, while the stronger pump drives the transition $|3\rangle \leftrightarrow |2\rangle$ almost resonantly. This leaves only the decay channel $|2\rangle \rightarrow |5\rangle$ open and therefore the population excited from the ground state $|1\rangle$ can be efficiently transferred to state $|5\rangle$. All the other decay channels have been closed by quantum coherence effects. This can be understood as a coherently-controlled version of a Raman process between states $|1\rangle - |2\rangle - |5\rangle$.  \Fref{fig:forpub1}$(b)$ shows that the transfer occurs only for nonzero rate $\gamma_{25}$; otherwise, the population stays in $|1\rangle$.

The explanation above, however, leaves open the question of why the transfer does not occur for a special case of $\delta_3 = \delta_4 = 0$ (see \fref{fig:forpub1}$(a)$). EIT  condition is satisfied, and this particular value of detuning cannot be singled out in the discussion above. To explain this, one needs to consider dressed states, i.e., the  eigenstates of the Hamiltonian of \Eref{eq:hambare}.

As we have seen, the population transfer from $|1\rangle $ to $|5\rangle $ takes place thanks to the decay channel $|2\rangle \to |5\rangle $ appearing in equations~(\ref{eq:rho22}) and~(\ref{eq:rho55}) and which is described by the operator $\gamma_{25}|2\rangle\langle5|$. Expressing this operator in the dressed basis is therefore an important step in search for the full and feasible explanation of this effect. In general, the transformation from atomic bare states to dressed states is $|e_i\rangle = \sum_j  U_{ij}|j\rangle$, where $|e_i\rangle$ denotes the dressed state and $|j\rangle$ denote atomic energy state, for $i,j = 1,\ldots,5$. Then, $|2\rangle\langle5| = \sum_{i,j} U^*_{i2}U_{j5} |e_i\rangle\langle e_j|$.

\begin{figure}[t]
\epsfxsize=12cm
\begin{center}
\epsfbox{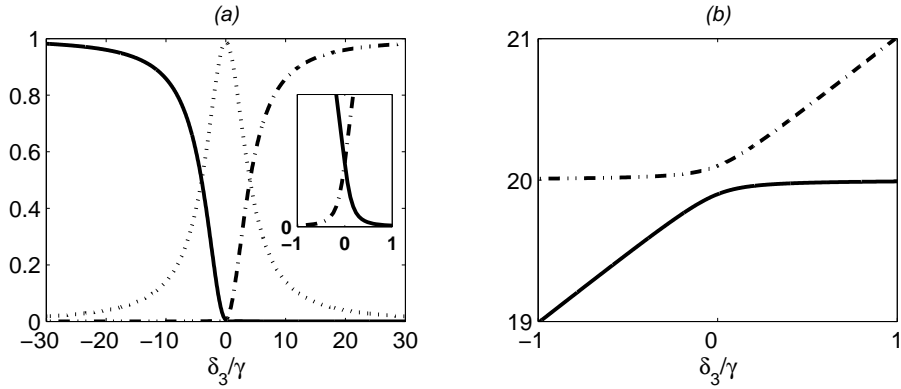}
\end{center}
\caption{\label{fig:dressab} $(a)$ Populations of the dressed states $|e_0\rangle$  (with eigenvalue $\epsilon_0=0$) (\dotted), $|e_3\rangle$ (eigenvalue $\epsilon_3$) (\chain) and $|e_4\rangle$ (eigenvalue $\epsilon_4$) (\full), for parameters as in \fref{fig:forpub1}$(a)$. The inset shows enlargement of the area around origin, with the exchange between the two dressed states. $(b)$ Eigenvalues $\epsilon_{3,4}$ (in units of $\gamma$) of the two dressed states as in $(a)$.}
\end{figure}

In the parameter region we are considering, i.e., $\delta_1 \gg \delta_2$, $\Omega_2 > \Omega_1 > \Omega_3,\Omega_4$, $\delta_3 \simeq \delta_4$ (see the caption of \fref{fig:forpub1} for example), the dark dressed state corresponding to the zero  eigenvalue $\epsilon_0=0$ is $|e_0\rangle \approx |1\rangle$; then there are two eigenvalues $\epsilon_{1,2} \approx \delta_1$ whose corresponding dressed states are not important here. The last two eigenvalues $\epsilon_3$ and $\epsilon_4$ and their corresponding dressed states $|e_{3,4}\rangle$ instead depend upon the explicit value of $\delta_3 = \delta_4$.
When $\delta_3 = \delta_4 \ll 0$, $\epsilon_3 \approx \delta_1$ and $\epsilon_4 \approx \delta_1+\delta_3$, while when $\delta_3 = \delta_4 \gg 0$ we have the reverse, i.e., $\epsilon_3 \approx \delta_1+\delta_3$ and $\epsilon_4 \approx \delta_1$. When $\delta_3 = \delta_4 \approx 0$ instead, $\epsilon_{3,4} \approx \delta_1 \pm O(\Omega_i)$ (see \fref{fig:dressab}$(b)$, which is obtained by numerical diagonalization and showing excellent agreement with these approximate expressions). In correspondence with this switching of eigenvalues also the expressions of the associated dressed states correspondingly switch: for $\delta_3 = \delta_4 \gg 0$, $|e_3\rangle \approx |5\rangle$ and $|e_4\rangle$ is dominated by $|2\rangle$ and $|4\rangle$, while when $\delta_3 = \delta_4 \ll 0$, states $|e_3\rangle$ and $|e_4\rangle$ switch roles. This is confirmed in \fref{fig:dressab}$(a)$, where the steady-state populations of the dressed states $|e_j\rangle $ $(j=1,3,4)$ obtained from the numerical solution of the Bloch equations, are shown.

The matrix of coefficients $U^*_{i2}U_{j5}$ of the expansion of $|2\rangle\langle5|$ in terms of dressed states is a sparse matrix. In the limit $\delta_3 = \delta_4 \gg 0$, only the term $|e_0\rangle\langle e_3|$ [and its Hermitian conjugate (H.c.)] survive. As expected by symmetry, in the opposite limit $\delta_3 = \delta_4 \ll 0$, only the term $|e_0\rangle\langle e_4|$ (and its H.c.) survive. At $\delta_3 = \delta_4 \approx 0$, both $|e_3\rangle$ and $|e_4\rangle$ are decoupled from $|e_0\rangle$, but coupled to each other, so that only the term  $|e_3\rangle\langle e_4|$ (and its H.c.) survive. This fact well explains the behavior of the steady-state populations as a function of the detuning $\delta_3=\delta_4$ and the corresponding possibility to realize an adiabatic population transfer: the decay term $\gamma_{25}$ couples $|1\rangle$ and $|5\rangle$ only for large $|\delta_{3,4}|$, but uncouples them for vanishing detunings.

\begin{figure}
\epsfxsize=12cm
\begin{center}
\epsfbox{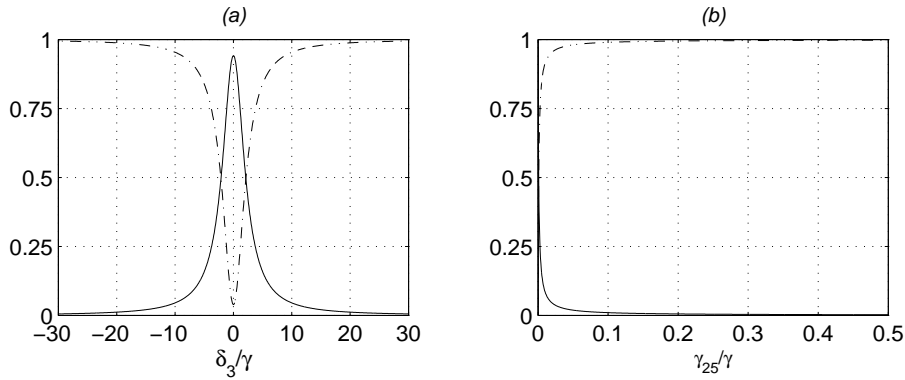}
\end{center}
\caption{\label{fig:forpub2}$(a)$ Populations $\rho_{33}$ (\full) and $\rho_{55}$ (\chain) for the detunings $\delta_{1,2}$ exchanged: $\delta_1 = 0$ and $\delta_2 = 20\gamma$. All the other parameters are the same as in \fref{fig:forpub1}$(a)$. In $(b)$, the same populations are shown for varying $\gamma_{25}$, and fixed $\delta_3 = \delta_4 = 20\gamma$ and other parameters as in $(a)$.}
\end{figure}

We also mention that basically the same mechanism can be used when atomic detunings  $\delta_1$ and $\delta_2$ exchange values. In this case, however, population transfer happens between $|3\rangle$ and $|5\rangle$, as shown in \fref{fig:forpub2}. The entire analysis outlined above can now be repeated, with the exchange $|1\rangle \leftrightarrow |3\rangle$ and $\delta_1 \leftrightarrow -\delta_2$.

Finally, it is obvious from the dressed states analysis that minor adjustments of the  M-scheme of \fref{fig:mscheme} produce another situation in which a population transfer can be achieved efficiently. Namely, if the laser of Rabi frequency $\Omega_2$ drives the transition $|2\rangle \leftrightarrow |5\rangle$, leaving transition $|2\rangle  \leftrightarrow |3\rangle$ uncoupled to light fields, the population transition similar to the one described above can be facilitated between levels $|3\rangle$ and $|5\rangle$. In this case, however, it is the cross-decay rate $\gamma_{14}$ that plays the role played by $\gamma_{25}$ in the initial M-scheme.

\section{Conclusion \label{sec:conclusion}}

In this article, we have presented a robust scheme for adiabatic population transfer between Zeeman-split ground states in a five-level atomic system. The scheme is well-suited for implementation with Rydberg atoms, but it can also be implemented in other media. It was shown that the transfer can be understood as a specific form of a Raman process, where numerous decay channels are simultaneously closed by quantum coherence effects. Only an incoherent decay channel is left open, realizing in this way the population transfer. The scheme operates in a steady state regime, so the need for a specific tailoring and sequence of a time dependent pulses is removed. Since population transfer is realized only thanks to a cross-decay between two atomic levels, we note that the present phenomenon cannot be described in terms of amplitude equations, as the crucial term $\gamma_{25}\rho_{22}$ in  \Eref{eq:rho55} cannot be obtained by the usual phenomenological generalization of Schr\"{o}dinger equation by inclusion of loss terms.

\ack

We acknowledge discussions with E. Arimondo. We also acknowledge financial support from the MIUR (PRIN 2002 \emph{Quantum Coherent Phenomena in Nonlinear Optics}).

\end{document}